\documentclass[aps,pra, twocolumn, noshowpacs, floatfix]{revtex4}
\usepackage{}

\usepackage{amsfonts}
\usepackage{amssymb}
\usepackage{graphicx}
\usepackage{amsmath}
\usepackage[english]{babel}
\usepackage{color}

\begin{document}

\title{Dynamics of dissipative Landau-Zener transitions}

\author{Zhongkai Huang and Yang Zhao\footnote{Electronic address:~\url{YZhao@ntu.edu.sg}}}
\affiliation{Division of Materials Science, Nanyang Technological University, Singapore 639798, Singapore}

\begin{abstract}
A non-perturbative treatment, the Dirac-Frenkel time-dependent variation is employed to examine dynamics of the Landau-Zener model with both diagonal and off-diagonal qubit-bath coupling using the multiple Davydov trial states. It is shown that steady-state transition probabilities agree with analytical predictions at long times. Landau-Zener dynamics at intermediate times
is little affected by diagonal coupling, and is found to be determined by off-diagonal coupling and tunneling strength between two diabatic states. We investigate effects of bath spectral densities, coupling strengths and interaction angles on Laudau-Zener dynamics. Thanks to the multiple Davydov trial states, detailed boson dynamics can also be analyzed in Landau-Zener transitions. Results presented here may help provide guiding principles to manipulate the Laudau-Zener transitions in circuit QED architectures by tuning off-diagonal coupling and tunneling strength.

\end{abstract}
\date{\today}

\maketitle

\section{introduction}
The Landau-Zener (LZ) transition comes into play when the energy difference between two diabatic states is swept through an avoided level crossing. Its final transition probability was calculated by Landau and Zener in $1932$ \cite{zener_1932, landau_1932}. As one of the most fundamental phenomena in quantum dynamics, the LZ transition plays an important role in a variety of fields, including atomic
and molecular physics \cite{thiel_1990, lipert_1988, xie_2017}, quantum optics \cite{bouw_1995}, solid state physics \cite{Wernsdorfer_2000}, chemical physics \cite{zhu_1997}, and quantum information science \cite{fuchs_2011}. The list of physical systems dominated by the LZ transition grows and interest in the LZ transition has been renewed recently due to its various new applications \cite{Onuchic_1988, petta_2010}, such as a nitrogen-vacancy center spin in isotopically purified diamond \cite{du_2014}, a microwave driven superconducting qubit coupled to a two level system, \cite{sun_2015} and a spin-orbit-coupled Bose-Einstein condensate \cite{olson_2014}.

In particular, advances in circuit quantum electrodynamics (QED) devices make them promising candidates for exploration of the LZ transitions due to their potential scalability and tunable parameters over a broad range \cite{saito_2006, oliver_2005, niemcyzk_2010}.
Circuit QED is the realization of cavity QED in superconducting quantum circuits. A superconducting flux qubit coupled to a quantum interference device \cite{chiorescu_2004} has been fabricated by Chiorescu {\it et al.}, as well as a charge qubit coupled to a transmission line resonator by Wallraff {\it et al.} \cite{wallraff_2004}. These developments have paved the way to study the LZ transitions because the energy difference between the two diabatic states has been allowed to be tuned by external fields \cite{wubs_2006}. Recent measurements of the LZ transitions have been reported on an individual flux qubit within a multiqubit superconducting chip, in which qubits are compound Josephson-junction radio frequency superconducting quantum interference device (SQUID) qubits \cite{Johansson_2009}.

In any physical realization, a quantum two-state system will be affected by its environment, which may alter the effective interaction between the two energy levels of the system. For a realistic study of a qubit manipulation via the LZ transitions, the influence of its environment is an important issue because a qubit is never completely isolated. Effects of dissipation have been studied in $1989$ by Ao {\it et al.}, using time-dependent perturbation theory, only giving the LZ transition probabilities at long times in the fast and slow sweeping limit \cite{ao_1989}.
H\"anggi and coworkers have studied the LZ transitions in a qubit coupled to a bath at zero temperature  and LZ dynamics with the master equation method \cite{saito_2007}.
Effects of temperature on the LZ transitions have been explored in a dissipative environment using the quasiadiabatic propagator path integral method and the non-equilibrium Bloch equations, providing only dependence of transition probability at long times on sweeping velocities \cite{nalbach_2009, nalbach_2014, nalbach_2015, dodin_2014}. Nalbach {\it et al.} have further studied the influence of thermal environment on a harmonically driven quantum two-state system through avoided crossings and proposed a novel rocking ratchet based on electronic double quantum dots \cite{nalbach_2017}. So far most attention has been paid to the transition probabilities in the steady states, where the energy difference of the two diabatic states is much larger than the bandwidth of the bosonic bath \cite{ashhab_2014}. However, more understanding of LZ dynamics at intermediate times is needed. 
This is a time range in which the transitions have not fully taken place and the energy difference of the two diabatic states is still within the bath's bandwidth \cite{orth_2010}. Specifically, dependence of LZ dynamics on the bath frequency and the types of bath spectral densities is still elusive.

Recently, high-quality fabrication techniques and physically large shunt capacitors have been developed to reduce densities and electric participation of defects at various metal and substrate interfaces, leading to rapid progresses in performance and manipulation of the flux qubit and its environment \cite{paik_2011}. An Ohmic type spectral density can be used to describe the qubit-bath coupling in various devices like a superconducting circuit consisting of a transmon qubit suspended on top of a microwave guide \cite{ripoll_2015}, a superconducting qubit interacting with an array of coupled transmission line resonators \cite{liu_2016}, and a fabricated circuit QED architecture that contains a capacitively shunted flux qubit coupled capacitively to a planar transmission line resonator \cite{yan_2015}. Egger {\it et al.} showed that a sub-Ohmic type spectral density can characterize the qubit-bath coupling in a multimode circuit QED setup with hybrid metamaterial transmission lines \cite{egger_2013}. Super-Ohmic type spectral densities have been applied to characterize the flux noise on multiple flux qubits, especially when scaling up to large numbers of qubits, as was stated by Storcz {\it et al.} \cite{storcz_2005, whitney_2011}. Nalbach et al. have uncovered that super-Ohmic fluctuations are the main relaxation channel for a detuned double quantum dot when the dot is driven by external voltage pulses \cite{nalbach_2013_prb}. When a superconducting persistent-current qubit is exposed to an underdamped SQUID¡¯s nvironment, Lorentzian spectral densities have usually been found \cite{tian_2002, zhe_2016}.

Dynamics of the LZ transitions at the intermediate times is influenced by the dissipative environment. Roles of the environment include fluctuations of energies of diabatic states, denoted by diagonal coupling, and environment-induced transitions between diabatic states, expressed by off-diagonal coupling. In the presence of only diagonal coupling, dynamics of the LZ transitions have been studied by Orth {\it et al.}, using a stochastic Schr\"odinger equation \cite{orth_2010, orth_2013}.
Off-diagonal coupling has been demonstrated to exist in a number of experiments, such as in a superconducting charge qubit coupled to an on-chip microwave resonator in the strong coupling regime \cite{wallraff_2004}, in a three-dimensional circuit QED architecture \cite{paik_2011}, a circuit QED device with seven qubits \cite{houck_2008}, and in a circuit QED implementation with a time-dependent transverse magnetic field \cite{viehmann_2013}. However, effects of off-diagonal coupling on LZ dynamics have not been well investigated. Recently, the multiple Davydov D$_2$ {\it Ansatz} has been developed to accurately treat dynamics of the generalized Holstein model with simultaneous diagonal and off-diagonal coupling \cite{zhou2015polaron, huang_2017}. Influences of off-diagonal coupling have also been probed in the intramolecular singlet fission model using our variational approach \cite{huang_SF_2017}.

In this work, we investigate impacts of diagonal and off-diagonal qubit-bath coupling on the standard LZ model using the multi-$\rm D_2$ {\it Ansatz} with the Dirac-Frenkel variational principle. The converged results by the employed method agree with those from other methods. In addition, calculated probabilities in the steady states concur with analytical predictions at zero temperature, further justifying the validity of our method.

The remainder of the paper is structured as follows. In Sec.~\ref{methodology}, we present the Hamiltonian and our trial
wave function, the multi-$\rm D_2$ {\it Ansatz}. In Sec.~\ref{Coupling to single bath mode}, a qubit coupled to a circuit oscillator is stuided. In Sec.~\ref{different spectral density}, the influence of bath spectral densities on the LZ transitions is investigated. Finally, effects of coupling strengths and interaction angles on LZ dynamics are examined in Sec.~\ref{different alpha and theta}. Conclusions drawn are in Sec.~\ref{Conclusions}.

\section{methodology}
\label{methodology}
\subsection{Model}
\label{Model}
The total Hamiltonian of a driven two-level system interacting with a bosonic bath is given by
\begin{eqnarray}
\hat{H}=\hat{H}_{S}+\hat{H}_{B}+\hat{H}_{SB}
\label{Hamiltonian}
\end{eqnarray}
where the system Hamiltonian is the standard LZ Hamiltonian for an isolated two-level system, {i.e}, $\hat{H}_{S}=\hat{H}_{LZ}$, with
\begin{eqnarray}
\hat{H}_{LZ}=\frac{vt}{2}\sigma_{z}+\frac{\Delta}{2}\sigma_{x}
\label{Hamiltonian_LZ}
\end{eqnarray}
where $\sigma_{x}$ and $\sigma_{z}$ are the Pauli matrices. The states, $\left|\uparrow\right\rangle$ and $\left|\downarrow\right\rangle$, are eigenstates of the qubit Hamiltonian ${\frac{vt}{2}\sigma_{z}}$. Energy difference between the diabatic states $vt$ varies linearly with time (with level-crossing speed $v>0$). Tunneling strength $\Delta$ represents intrinsic interactions between the two diabatic states, and induces the transitions.

To consider the Landan-Zener transition in the presence of an environment, we model a bosonic bath of $N$ quantum harmonic oscillators by the Hamiltonian $\hat{H}_{B}$ and the qubit-bath coupling by the Hamiltonian $\hat{H}_{SB}$ \cite{wubs_2006},
\begin{eqnarray}
\hat{H}_{B}&=&\sum_{q=1}^{N}\hbar\omega_{q}\hat{b}_{q}^{\dagger}\hat{b}_{q}\nonumber\\
\hat{H}_{SB}&=&\sum_{q=1}^{N}\frac{\gamma_{q}}{2}\left(\cos\theta_{q}\sigma_{z}+\sin\theta_{q}\sigma_{x}\right)(\hat{b}_{q}^{\dagger}+\hat{b}_{q})
\end{eqnarray}
where $\hbar=1$ is assumed throughout, $\omega_q$ indicates the frequency of the $q$-th mode of the bath with creation (annihilation) operator $\hat{b}_{q}^{\dagger}(\hat{b}_{q})$. $\gamma_i$ and $\theta_i$ are the qubit-oscillator coupling and the interaction angle, respectively. The effect of the bosonic bath is to change the energies of the qubit via the diagonal coupling ($\sigma_z$) and to induce transitions between the levels of the qubit via the off-diagonal coupling ($\sigma_x$)

The environment and its coupling to the system are characterized by a spectral density function,
\begin{eqnarray}
J\left(\omega\right)=\sum_{q}\gamma_{q}^{2}\delta\left(\omega-\omega_{q}\right)=2\alpha\omega_{c}^{1-s}\omega^{s}e^{-\omega/\omega_{c}}
\label{spectral}
\end{eqnarray}
where $\alpha$ is the dimensionless coupling strength, $\omega_{c}$ denotes the cutoff frequency, and $s$ determines the dependence of $J\left(\omega\right)$ on the bath frequency $\omega$. The bosonic Ohmic bath is specified by $s=1$, and $s<1(s>1)$ denotes the sub-Ohmic (super-Ohmic) bath \cite{whitney_2011}.
The effect of spectral density of Lorentzain line shape on LZ dynamics will be studied in future work.

\subsection{The Multi-D$_2$ state}
\label{Multiple Davydov trial state}
The multiple Davydov trial states with multiplicity $M$ are essentially $M$ copies of the corresponding single Davydov {\it Ansatz} \cite{zh_12,zh_97}. They were developed to investigate the polaron model \cite{zhou2015polaron, huang_2017, huang_2017_off} and the spin-boson model \cite{huang_SF_2017} following the Dirac-Frenkel variational principle. In the two-level system, one of the multiple Davydov trial states, the multi-${\rm D}_2$ {\it Ansatz} with multiplicity $M$, can be constructed as
\begin{eqnarray}\label{D2_state}
&&\left|D_{2}^{M}\right\rangle =\sum_{i=1}^{M}\left\{ A_{i}(t)\left|\uparrow\right\rangle \exp{\left[\sum_{q=1}^{N}f_{iq}(t)\hat{b}_{q}^{\dagger}-H.c.\right]}\left|0\right\rangle \right\}\nonumber\\
&&+\sum_{i=1}^{M}\left\{B_{i}(t)\left|\downarrow\right\rangle \exp{\left[\sum_{q=1}^{N}f_{iq}(t)\hat{b}_{q}^{\dagger}-H.c.\right]}\left|0\right\rangle \right\},
\end{eqnarray}
where $H.c.$ denotes the Hermitian conjugate, and $\left|0\right\rangle$ is the vacuum state of the bosonic bath. $A_{i}$ and $B_{i}$ are time-dependent variational parameter for the amplitudes in states $\left|\uparrow\right\rangle$ and $\left|\downarrow\right\rangle$, respectively, and $f_{iq}(t)$ are the bosonic displacements, where $i$ and $q$ label the $i$-th coherent superposition state and $q$-th effective bath mode, respectively. If $M=1$, the multi-$D_2$ {\it Ansatz} is restored to the usual Davydov $\rm D_2$ trial state.

Equations of motion of the variational
parameters $u_{i}=$$A_{i}, B_{i}$ and $f_{iq}$ are then derived by adopting the Dirac-Frenkel variational principle,
\begin{eqnarray}\label{eq:eom1}
\frac{d}{dt}\left(\frac{\partial L}{\partial\dot{u_{i}^{\ast}}}\right)-\frac{\partial L}{\partial u_{i}^{\ast}} & = & 0.
\end{eqnarray}
For the multi-$\rm D_2$ {\it Ansatz}, the Lagrangian $L_2$ is formulated as
\begin{eqnarray}
L_2 & = & \langle {\rm D}^M_2(t)|\frac{i\hbar}{2}\frac{\overleftrightarrow{\partial}}{\partial t}- \hat{H}|{\rm D}^M_2(t)\rangle \nonumber \\
& = & \frac{i\hbar}{2}\left[ \langle {\rm D}^M_2(t)|\frac{\overrightarrow{\partial}}{\partial t}|{\rm D}^M_2(t)\rangle - \langle {\rm
D}^M_2(t)|\frac{\overleftarrow{\partial}}{\partial t}|{\rm D}^M_2(t)\rangle \right] \nonumber \\
&-& \langle {\rm D}^M_2(t)|\hat{H}|{\rm D}^M_2(t)\rangle.
\label{Lagrangian_2}
\end{eqnarray}
Details of the Lagrangian, equations of motion, and initial conditions are given in Appendix \ref{Equations of Motion}.

\section{Results and discussion}
\label{Numerical results and discussions}

\subsection{A qubit coupled to a single mode}
\label{Coupling to single bath mode}
\begin{figure}[tbp]
\centering
\includegraphics[scale=0.425]{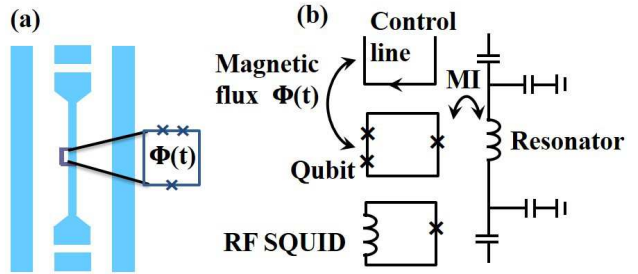}
\caption{(a) Schematic diagram of a typical coplanar waveguide resonator with a qubit placed between the center conductor and the ground plane of the waveguide. (b) Sketch of the superconducting qubit coupled to the coplanar transmission line resonator. MI denotes the mutual inductance between the qubit and resonator. The control line supplies the time-dependent magnetic flux $\Phi(t)$ threading the qubit loop.}
\label{sketchQED}
\end{figure}

The LZ transitions can occur in a qubit that is coupled to a circuit oscillator in a QED device \cite{chiorescu_2004, wallraff_2004}. Fig.~\ref{sketchQED} displays the schematic diagram of a superconducting qubit coupled to a coplanar transmission line resonator. The control line in Fig.~\ref{sketchQED}(b) supplies the time-dependent magnetic flux $\Phi(t)$ threading a persistent current qubit loop, which contains three junctions. After manipulations of the qubit, the state is detected by a SQUID, which consists of a single Josephson junction in a superconducting loop \cite{lind_2007}. By tuning the external magnetic flux $\Phi(t)$ threading the qubit loop, the energy level separation can vary linearly with a level-crossing speed $v$. The resonator can represent a single mode of harmonic oscillator, and is coupled to the qubit. Then this qubit-oscillator setup can simply be modeled by a Hamiltonian
\begin{eqnarray}
\hat{H}=\frac{vt}{2}\sigma_{z}+\frac{\Delta}{2}\sigma_{x}+\hbar\omega\hat{b}^{\dagger}\hat{b}+\frac{\gamma}{2}\sigma_{x}(\hat{b}^{\dagger}+\hat{b}),
\label{Hamiltonian_LZ_single}
\end{eqnarray}
which can be obtained from the Hamiltonian~(\ref{Hamiltonian}) if the number of modes is set to one ($N=1$). When the first term in Eq.~(\ref{Hamiltonian_LZ_single}) is replaced by time-independent energy bias, the Hamiltonian is reduced to be the Rabi model, a paradigmatic construct of a two-level system coupled to a single bosonic mode derived from an atom in an applied electric field. A conventional rotating-wave approximation has often been adopted to treat the Rabi model \cite{zhang_2015}.

Transitions between two diabatic states can result from direct tunneling strength or indirect off-diagonal coupling to the oscillator. The physical quantity of interest includes the probability that the qubit flipped from the initial state $\left|\uparrow\right\rangle$ to $\left|\downarrow\right\rangle$, {i.e.}, $P_{\uparrow\rightarrow\downarrow}\left(\infty\right)=1-P_{\uparrow\rightarrow\uparrow}\left(\infty\right)$. Concerning the tunneling strength between the two diabatic states, the final transition probability through avoided level crossing point is given by the familiar Landu-Zener formula $P_{LZ}=1-\exp\left(\frac{-\pi\Delta^{2}}{2\hbar\left|v\right|}\right)$ \cite{zener_1932, landau_1932, witting_2005, chichinin_2013, ho_2014}. With respect to the indirect off-diagonal coupling to the single bath mode, the transition probability is proposed as $P_{\uparrow\rightarrow\downarrow}\left(\infty\right)=1-\exp\left(\frac{-\pi\gamma^{2}}{2\hbar\left|v\right|}\right)$ at zero temperature \cite{wubs_2006, saito_2006}. In this work, we have studied the combined effect of the direct tunneling strength between the two diabatic states and indirect off-diagonal coupling to the single bath mode. Niemczyk {\it et al.} \cite{niemcyzk_2010} in a recently developed circuit QED device showed the breakdown of the widely used rotating-wave approximation and master equation method due to the existence of strong qubit-bath coupling \cite{sun_2012}.

\begin{figure}[tbp]
\centering
\includegraphics[scale=0.425]{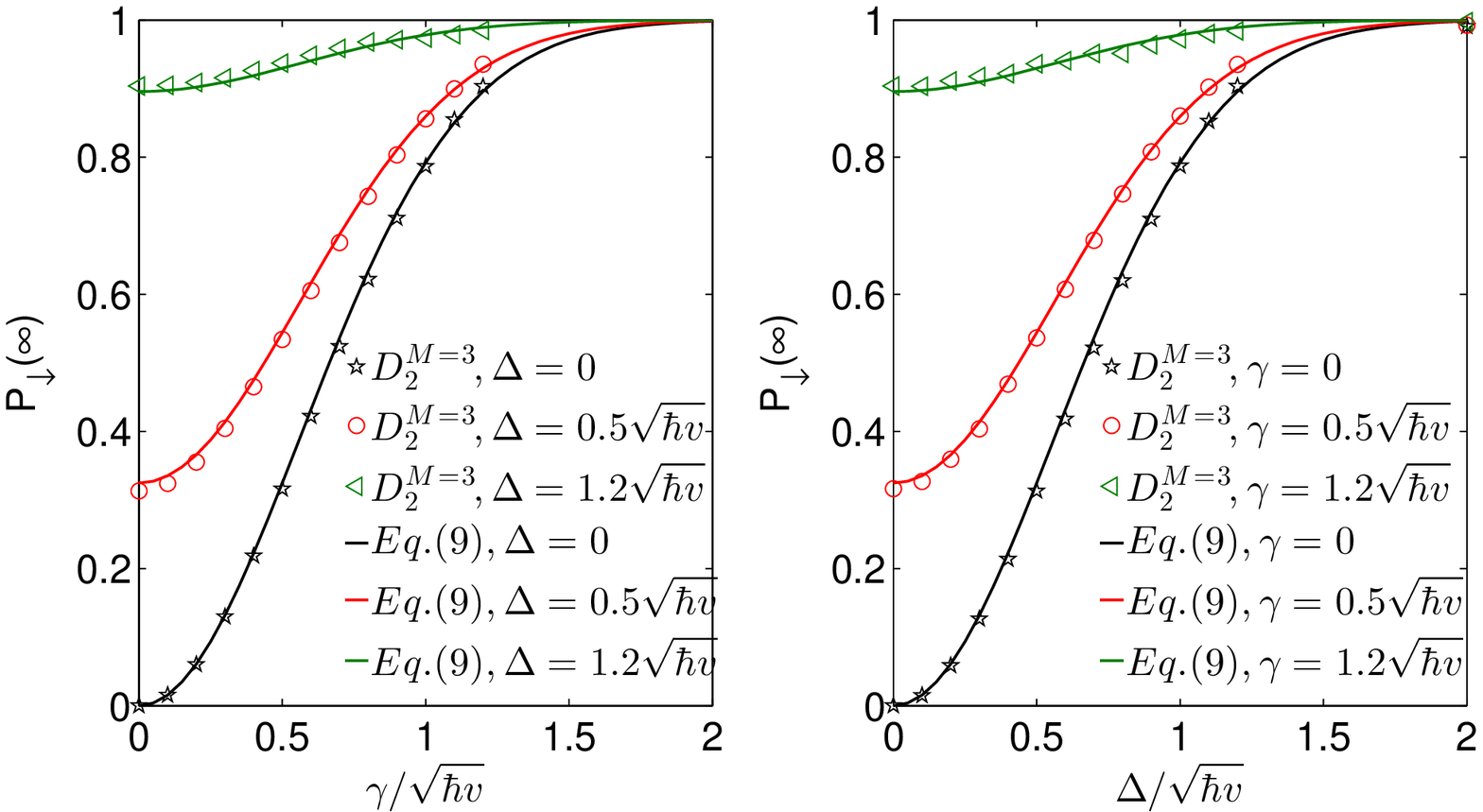}
\caption{ (a) Final transition probability $P_{\uparrow\rightarrow\downarrow}\left(\infty\right)$ as a function of off-diagonal coupling strength $\gamma/\sqrt{\hbar v}$ with fixed tunneling strength $\Delta=0$, $0.5\sqrt{\hbar v}$ and $1.2\sqrt{\hbar v}$. (b) $P_{\uparrow\rightarrow\downarrow}\left(\infty\right)$ as a function of tunneling strength $\Delta/\sqrt{\hbar v}$ for different off-diagonal coupling strengths $\gamma=0$, $0.5\sqrt{\hbar v}$ and $1.2\sqrt{\hbar v}$. The oscillator frequency $\omega$ is set to $10\sqrt{v/\hbar}$.}
\label{exactdifDdifC}
\end{figure}

Using the time perturbation theory \cite{saito_2007}, we obtain
\begin{eqnarray}
P_{\uparrow\rightarrow\downarrow}\left(\infty\right)=1-\exp\left[\frac{-\pi(\Delta^{2}+\gamma^{2})}{2\hbar\left|v\right|}\right] \label{final_single}
\end{eqnarray}
It has been shown that this formula can provide exact final transition probabilities for the whole parameter regime at zero temperature \cite{saito_2006, wubs_2006}. As shown in Fig.~\ref{exactdifDdifC}, $P_{\uparrow\rightarrow\downarrow}\left(\infty\right)$ calculations from the multi-D$_2$ {\it Ansatz} with a sufficiently large multiplicity $M$ agrees with the analytical predictions of Eq.~(\ref{final_single}) for various off-diagonal coupling strengths $\gamma$ and tunneling strengths $\Delta$. This demonstrates the accuracy of our multi-$D_2$ {\it Anstaz} and we can numerically  provide the accurate final transition probabilities.

\begin{figure}[tbp]
\centering
\includegraphics[scale=0.45]{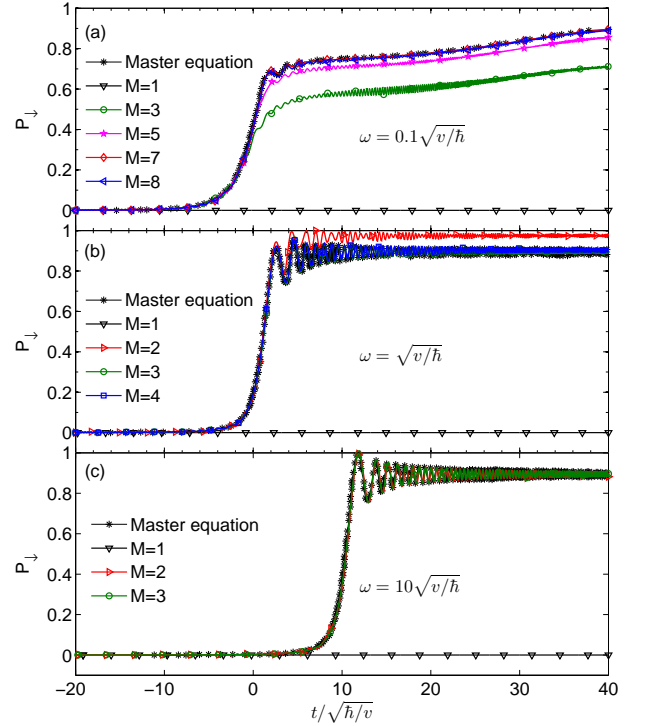}
\caption{Time evolution of transition probability calculated by the master equation method and the multi-$D_2$ {\it Ansatz}. Oscillator frequencies used are (a) $\omega=0.1 \sqrt{v/\hbar}$, (b) $\omega=\sqrt{v/\hbar}$ and (c) $\omega=10 \sqrt{v/\hbar}$. Other parameters are $\Delta=0$ and $\gamma=1.2\sqrt{\hbar v}$.}
\label{SI_converge_Omeaga}
\end{figure}

We here further justify the validity of the variational method by a comparison with the master equation method that yields exact results in the weak coupling regime. It is known that the multi-D${_2}$ {\it Ansatz}, a superposition of coherent states, can easily treat exciton dynamics in the strong coupling regime \cite{zhou2015polaron, huang_2017, huang_2017_off}. To reach an accurate description in the weak coupling cases, we have used a variety of multiplicities $M$ of the multi-D${_2}$ in the corresponding dynamical calculations. Fig.~\ref{SI_converge_Omeaga}(a), (b), and (c) display the time evolution of the transition probability with oscillator frequencies of $\omega=0.1\sqrt{v/\hbar}$, $\sqrt{v/\hbar}$, and $10\sqrt{v/\hbar}$, respectively. The multiplicity of the multi-D$_2$ {\it Ansatz} needed for convergence, as expected, decreases as the oscillator frequency increases if the coupling strength $\gamma$ stays constant. The converged results in each scenario concur with those extracted from Ref.~\cite{saito_2006} (black line with stars) using the master equation method, demonstrating that the multi-D$_2$ {\it Anstaz} can well describe the LZ dynamics at intermediate times when the qubit is coupled to the harmonic oscillator of a wide range of frequencies.

\begin{figure}[tbp]
\centering
\includegraphics[scale=0.48]{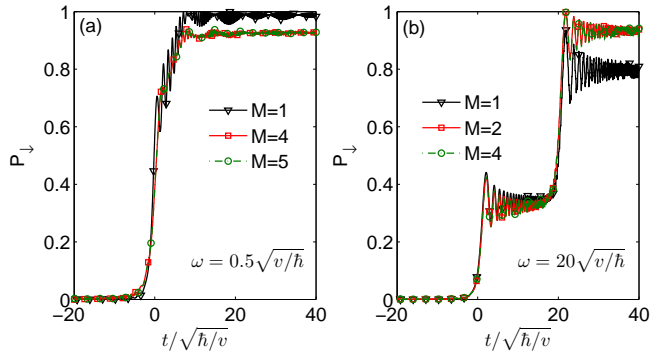}
\caption{ LZ dynamics with tunneling strength $\Delta=0.5\sqrt{\hbar v}$ and a off-diagonal coupling strength $\gamma=1.2\sqrt{\hbar v}$ for two oscillator frequencies (a) $\omega=0.5\sqrt{v/\hbar}$ and (b) $\omega=20\sqrt{v/\hbar}$.}
\label{converge}
\end{figure}

In order to gain insight into LZ dynamics at intermediate times, we also perform convergence tests for oscillator frequencies of $\omega=0.5\sqrt{v/\hbar}$ and $20\sqrt{v/\hbar}$, and results are shown in Figs.~\ref{converge}(a) and (b), respectively. In the absence (Fig.~\ref{SI_converge_Omeaga}) and the presence (Fig.~\ref{converge}) of the tunneling strength, it can be found that LZ dynamics at intermediate times strongly depends on the oscillator frequency $\omega$, while the steady-state population in $\left|\downarrow\right\rangle$, $P_{\downarrow}\left(\infty\right)$, are independent of $\omega$. In particular, the transition is temporally shifted from $t=0$ to $t=\hbar\omega/v$ due to the indirect off-diagonal coupling \cite{saito_2006}. Therefore the time shift for the case of $\omega=0.5\sqrt{v/\hbar}$ is minor compared to the time scale that is concerned, leading to the LZ transition of only one stage in Fig.~\ref{converge}(a). In contrast, $P_{\downarrow}(t)$ undergoes two stages of the LZ transitions in Fig.~\ref{converge}(b). The first transition stage is induced by direct tunneling strength between the two levels $\Delta=0.5\sqrt{\hbar v}$, named after the standard LZ transition, while the second transition stage results from the indirect off-diagonal coupling to the single oscillator mode with the frequency of $\omega=20\sqrt{v/\hbar}$.

\begin{figure}[tbp]
\centering
\includegraphics[scale=0.48]{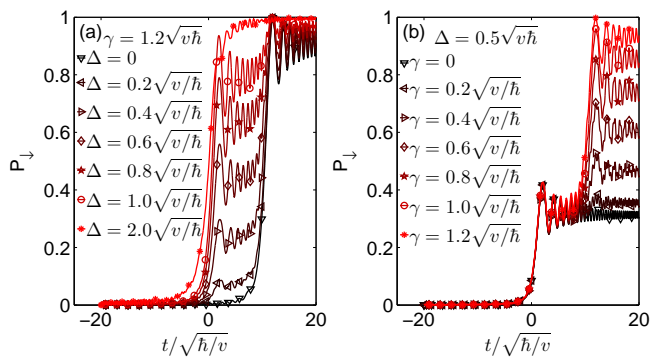}
\caption{ LZ dynamics (a) for six tunneling strengths $\Delta=0$, $0.2\sqrt{\hbar v}$, $0.4\sqrt{\hbar v}$, $0.6\sqrt{\hbar v}$, $0.8\sqrt{\hbar v}$, $1.0\sqrt{\hbar v},$ and $2.0\sqrt{\hbar v}$ with fixed $\gamma=1.2\sqrt{\hbar v}$ and (b) for different off-diagonal coupling strengths $\gamma=0$, $0.2\sqrt{\hbar v}$, $0.4\sqrt{\hbar v}$, $0.6\sqrt{\hbar v}$, $0.8\sqrt{\hbar v}$, $1.0\sqrt{\hbar v},$ and $1.2\sqrt{\hbar v}$ with certain $\Delta=0.5\sqrt{\hbar v}$. The oscillator frequency $\omega$ is set to $10\sqrt{v/\hbar}$.}
\label{difDdifC}
\end{figure}

Next, we  have investigated the dependence of LZ dynamics on the direct tunneling strength between the two diabatic states and indirect off-diagonal coupling to the single oscillator mode. For this simulation, the oscillator frequency of $\omega=10\sqrt{v/\hbar}$ has been used. As shown in Fig.~\ref{difDdifC}(a), by evenly changing the tunneling strength, the first plateau between the two stages of transitions can be tuned nonlinearly from zero to almost one, and the height of the second plateau varies from $0.89$ to $1$. As presented in Fig.~\ref{difDdifC}(b), the first plateau is kept around $0.32$ and the second plateau increases toward $1$ as the off-diagonal coupling strength increases. Results in this subsection offer the possibility to manipulate the quantum states of the qubit that is coupled to only one circuit oscillator in the circuit QED.

\subsection{Effect of the bath spectral density}
\label{different spectral density}

Recent developments in circuit QED setups have shown that qubits can couple to a bath of quantum harmonic oscillators also \cite{yan_2015, ripoll_2015, liu_2016}.
The qubit-bath coupling can be characterized by spectral densities of Ohmic type in a superconducting circuit consisting of a transmon qubit suspended on top of a microwave guide \cite{yan_2015}. Many theoretical efforts have also been devoted to study LZ transitions at long times in the dissipative environment in terms of Ohmic fluctuations \cite{nalbach_2010}. Spectrum densities of Sub-Ohmic type and super-Ohmic type can be realized in a multimode circuit QED setup with hybrid metamaterial transmission lines \cite{egger_2013} and in certain circuit QED setups with multiple flux qubits \cite{storcz_2005}, respectively. Thus effects of spectral densities and coupling strengths on LZ dynamics of these systems need to be addressed.

In this section, we have studied LZ dynamics using the spectral density of Eq.~(\ref{spectral}). We have assumed that all bath oscillators couple to the qubit with identical coupling angles $\theta_q=\theta$. We can have integrated spectral density $S=\sum_{q}\gamma_{q}^{2}=\frac{\hbar^2}{4\pi}\int_{0}^{\infty}d\omega J\left(\omega\right)=\frac{\hbar^2}{2\pi}\alpha\omega_{c}^{s+1}\Gamma\left(s+1\right)$
and total reorganization energy $E_{0}=\frac{\hbar}{4\pi}\int_{0}^{\infty}d\omega\frac{J\left(\omega\right)}{\omega}=\frac{\hbar}{2\pi}\alpha\omega_{c}^{s}\Gamma\left(s\right),$
where $\Gamma\left(x\right)$ is the Euler gamma function. Thus the final transition probability at zero temperature \cite{saito_2007} can be given as
\begin{eqnarray}
&&P_{\uparrow\rightarrow\downarrow}\left(\infty\right)=\nonumber\\
&&1-\exp\left[\frac{-\pi\left(\left|\Delta-\frac{1}{2}E_{0}\sin\left(2\theta\right)\right|^{2}+S\sin^{2}\theta\right)}{2\hbar v}\right]
\label{final_ohmic}.
\end{eqnarray}
When the first term in Eq.~(\ref{Hamiltonian_LZ}) is replaced by time-independent term of $\frac{\epsilon}{2}\sigma_z$, the Hamiltonian~(\ref{Hamiltonian}) becomes a spin-boson Hamiltonian. When system-bath coupling increases, a delocalization-localization transition can be found within the framework of the spin-boson model \cite{wang_2016}. For LZ problems, however, the system always reaches a steady state with a certain final transition probability because the energy difference between the two diabatic states will be so large that transitions between the two states are unlikely at long times.

\begin{figure}[tbp]
\centering
\includegraphics[scale=0.5]{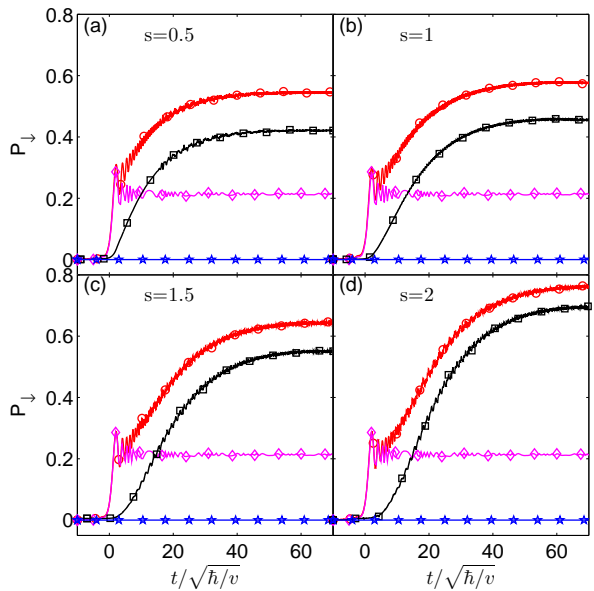}
\caption{Time evolution of transition probability for (a) a sub-Ohmic bath of $s=0.5$, (b) an Ohmic bath of $s=1$, and super-Ohmic bath of (c) $s=1.5$ and (d) $s=2$ is obtained from the $\rm D_2^{M=3}$ {\it Ansatz} with an identical  coupling strength $\alpha=0.002$. For each of the four $s$ values, four cases are shown: $\Delta=0.4\sqrt{\hbar v}, \theta=\pi/2$ (red line, circles), $\Delta=0.4\sqrt{\hbar v}, \theta=0$ (magenta line, diamonds), $\Delta=0, \theta=\pi/2$ (black line, squares), and $\Delta=0, \theta=0$ (blue line, pentagrams).}
\label{difspectral}
\end{figure}

As shown in Fig.~\ref{difspectral}, we compare the LZ dynamics of the sub-Ohmic, Ohmic and super-Ohmic bath with the same coupling strength $\alpha=0.002$. We have calculated the converged results of LZ dynamics for a qubit coupled to baths using the variational method. Spectral densities of the sub-Ohmic bath are computed using logarithmic discretization. For Ohmic and super-Ohmic bath we have used linear discretization \cite{wang_2016}. Cutoff frequency is given by $\omega_c=10\sqrt{v/\hbar}$. The roughness of the curves can be significantly reduced by using a large number of frequency modes ($N=80$ or greater). Details on the convergence tests are presented in Appendix \ref{convergence test}.

In Figs.~\ref{difspectral} (a) and (b) we have presented LZ dynamics for the sub-Ohmic bath ($s=0.5$) and the Ohmic bath ($s=1$), respectively. Figs.~\ref{difspectral}(c) and (d) depict time evolution of transition probabilities using the super-Ohmic bath with $s=1.5$ and $2$, respectively. When $\theta$ $=0$, there exists only one stage of the LZ transition near $t=0$ for nonzero tunneling strength. That is, in the presence of only diagonal coupling, LZ dynamics of $\Delta=0.4\sqrt{\hbar v}$ (magenta lines, diamonds) are almost identical in the four subplots, irrespective of the spectral densities. Further calculations with finite tunneling strengths have shown that there exists a one-stage LZ transition in general in the presence of diagonal coupling only.

When $\theta=\pi/2$, time evolution of the transition probability for $\Delta=0$ has a single stage of slow growth until it reaches its steady state. The converged probabilities and the convergence times are dependent on the spectral densities. This occurs because LZ dynamics is strongly dependent on the oscillator frequency $\omega$ for a qubit off-diagonally coupled to a single harmonic oscillator, as has been shown in Sec.~\ref{Coupling to single bath mode}. Fig.~\ref{difspectral} also depicts that the convergence time for large $s$ is significantly longer than that for smaller $s$, since spectral densities of large $s$ involve prominent contribution from high-frequency oscillators, and the convergence time in the single harmonic oscillator scenario is proportional to the oscillator frequency $\omega$. When $\Delta=0.4\sqrt{\hbar v}$, there are two stages of the LZ transitions in the presence of off-diagonal coupling. In the first stage, transition probability jumps up at $t=0$. In the second stage, it gradually reaches the steady state at the same convergence time as that of $\Delta=0$. Further calculations have shown that there exist the two-stage LZ transitions in general for all non-zero tunneling strengths in the presence of off-diagonal coupling. In addition, as expected, the converged transition probabilities obtained from our dynamics calculations agree with the corresponding steady-state transition probabilities from Eq.~(\ref{final_ohmic}).

\begin{figure}[tbp]
\centering
\includegraphics[scale=0.42]{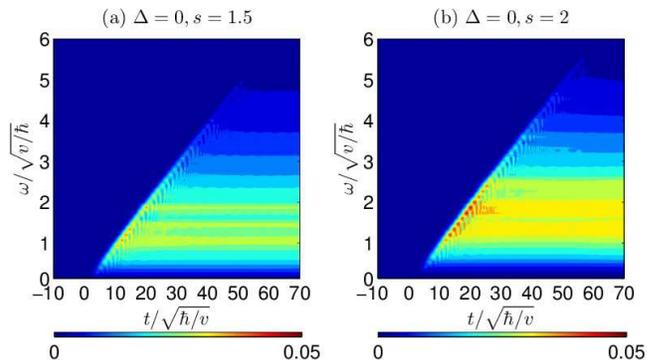}
\caption{Time evolution of the boson number using for (a) a super-Ohmic bath of $s=1.5$ and (b) $s=2$, in the presence of off-diagonal coupling only ($\theta=\pi/2$). Other parameters are $\Delta=0$ and $\alpha=0.002$.}
\label{Ph_pop_difs}
\end{figure}

To investigate the role of bosons in the LZ transitions, we have calculated the time evolution of the boson number $\langle\hat{b}_{q}^{\dagger}\hat{b}_{q}\rangle$ which is shown in Fig.~\ref{Ph_pop_difs}.
The initial boson number is set to be zero in our calculations. The bosons will be created after the transition takes place. If the qubit is only off-diagonally coupled to the single harmonic oscillator, the LZ transition would be temporally shifted from $t=0$ to $ t = \hbar\omega/v$, independent of the coupling strength \cite{saito_2006}. If qubit is off-diagonally coupled to multiple harmonic oscillators, the transition will then occur mainly after $t=0$ as there is a temporal shift of each frequency mode, as shown in Fig.~\ref{Ph_pop_difs}. Because the energy difference between the diabatic states varies linearly with time, the frequency of the bosons created via qubit-bath coupling also has the same time dependence, resulting in the upper edge of the triangle starting from $t=0$ in the $\omega-t$ plots. It can be found that very few bosons will be created for $t<0$, regardless of $s$ and coupling strengths. When a larger value of $s$ is used, more high-frequency bosons are created and this results in a larger steady-state probability for identical coupling strength. Also the time taken to create high-frequency mode bosons increases, which can be seen on comparison of  Figs.~\ref{Ph_pop_difs}(b) and Figs.~\ref{Ph_pop_difs}(a). This was expected from the convergence time taken to reach the steady states in Fig.~\ref{difspectral}(d) and (c).

If the energies corresponding to frequencies of the bath modes $\omega$ are high in comparison with the thermal energy $k_{B}T$, the oscillators are thermally inactive, thus LZ dynamics driven by the bath modes is temperature independent in a wide temperature range \cite{wallraff_2004, chiorescu_2004}. Therefore, the temperature can be set to be $T=0$ to reduce the numerical cost, although the inclusion of the temperature effect in the multiple Davydov {\it Ansatz} is straightforward by applying Monte Carlo importance sampling \cite{wang_2017}.

\subsection{Effects of coupling strength and interaction angle}
\label{different alpha and theta}

\begin{figure}[tbp]
\centering
\includegraphics[scale=0.42]{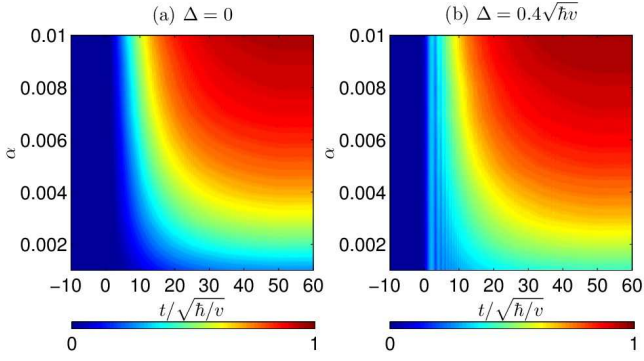}
\caption{Time evolution of transition probability for (a) $\Delta=0$ (b) $\Delta=0.4\sqrt{\hbar v}$ using an Ohmic bath with various coupling strengths $\alpha$, in the presence of off-diagonal coupling only ($\theta=\pi/2$).}
\label{population_t_alpha}
\end{figure}

Even though effects of various spectral densities have been discussed in Sec.~\ref{different spectral density}, we will focus on the Ohmic type in this subsection because of the recent progress in nanotechnology \cite{castellanos_2007, manucharyan_2009, chung_2009, pop_2010} which allows for feasible control of how Ohmic environments are coupled to the superconducting qubit \cite{cedraschi_2000, kopp_2007}. Figs.~\ref{population_t_alpha}(a) and (b) present the time evolution of transition probability as a function of coupling strength $\alpha$ for two values of tunneling strength, $\Delta=0$ and $0.4\sqrt{\hbar v}$, respectively. In this section we have considered the case for off-diagonal coupling ($\theta=\pi/2$) only. The calculated steady-state probabilities agree with  Eq.~(\ref{final_ohmic}), which predicts increases of the probabilities with the coupling strength.
While the coupling strengths for the left and right panels in Fig.~\ref{population_t_alpha} are the same, the steady-state probabilities of Fig.~\ref{population_t_alpha}(b) are larger than those of Fig.~\ref{population_t_alpha}(a), because the nonzero tunneling strength $\Delta=0.4\sqrt{\hbar v}$ gives rise to one more transition stage at $t=0$ compared to that of $\Delta=0$.

\begin{figure}[tbp]
\centering
\includegraphics[scale=0.42]{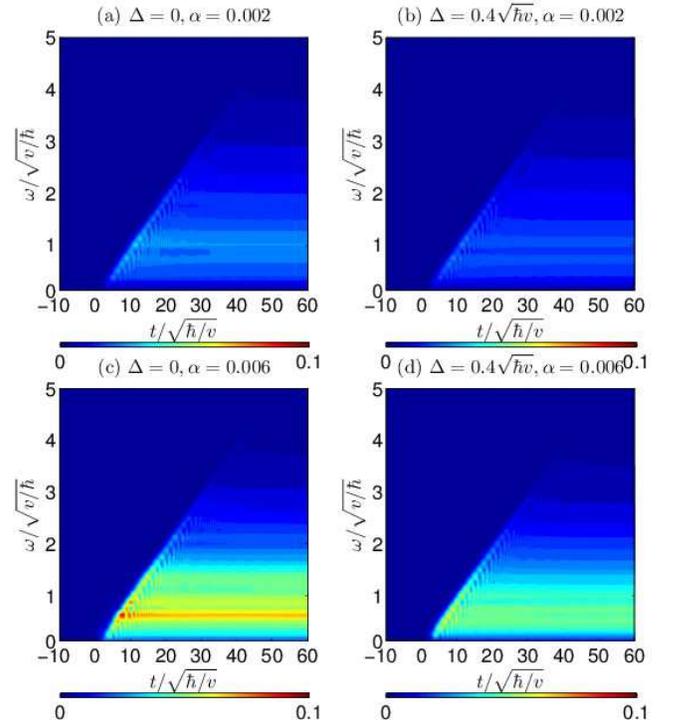}
\caption{Time evolution of the boson number using an Ohmic bath, in the presence of off-diagonal coupling only ($\theta=\pi/2$). The left column corresponds to $\Delta=0$, while the right column is for $\Delta=0.4\sqrt{\hbar v}$. The upper and lower panels correspond to coupling strength of $\alpha=0.002$ and $\alpha=0.006$.}
\label{Pop_ph_D0D04_difapha}
\end{figure}

The interplay between the circuit qubit and the bosons is characterized by boson dynamics as a function of $\omega$, as is shown in Fig.~\ref{Pop_ph_D0D04_difapha}. The boson number is initialized to zero. The upper and lower panels correspond to coupling strengths of $\alpha=0.002$ and $\alpha=0.006$, respectively. It was found that boson number becomes larger with stronger off-diagonal coupling. We then make a comparison between the left and right panels, in which the left column corresponds to the zero tunneling strength scenarios ($\Delta=0$) and the right column is for $\Delta=0.4\sqrt{\hbar v}$. If off-diagonal coupling strength is the same, more bosons are created for weaker tunneling scenarios, though we have larger steady-state transition probabilities for larger tunneling strength cases.

\begin{figure}[tbp]
\centering
\includegraphics[scale=0.42]{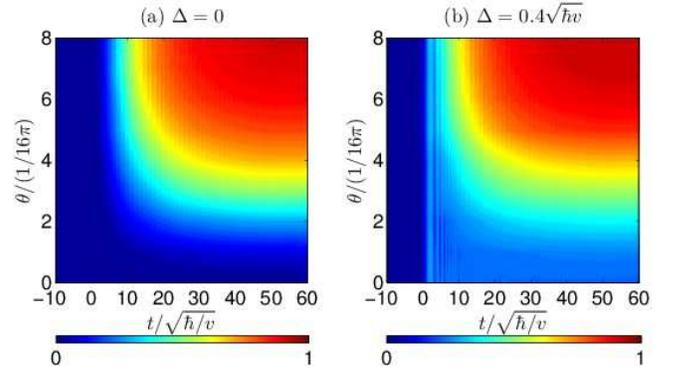}
\caption{Time evolution of transition probability for (a) $\Delta=0$ (b) $\Delta=0.4\sqrt{\hbar v}$ (to be replaced with more accurate calculations) using an Ohmic bath with various interaction angles $\theta$. The coupling strength $\alpha=0.008$ is set.}
\label{population_t_theta}
\end{figure}

\begin{figure}[tbp]
\centering
\includegraphics[scale=0.42]{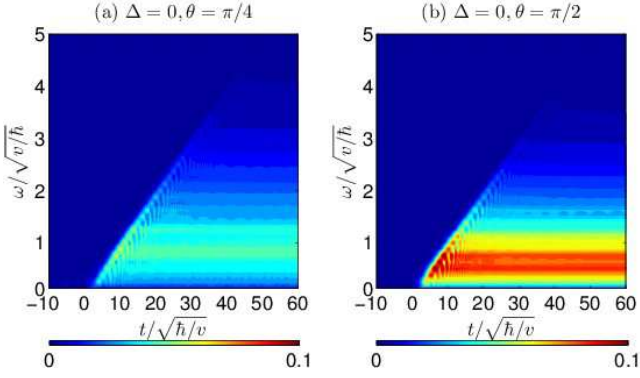}
\caption{Time evolution of the boson number using an Ohmic bath for interaction angles of (a) $\theta=\pi/4$ and (b) $\theta=\pi/2$. The tunneling strength $\Delta=0$ and coupling strength $\alpha=0.008$ are set.}
\label{Pop_Ph_D0_diftheta}
\end{figure}

Figs.~\ref{population_t_theta}(a) and (b) present the time evolution of the transition probability as a function of the interaction angle $\theta$  for $\Delta=0$ and $\Delta=0.4\sqrt{\hbar v}$, respectively. The interaction angle $\theta$ of interest ranges from $0$ to $\pi/2$. We have only considered coupling strength of $\alpha=0.008$ in this section. In the absence and presence of the tunneling strength, the LZ transitions shows one stage and two stages, respectively. The transition probabilities $P_{\downarrow}(t)$ for $t>0$ increase with interaction angles $\theta$, since the larger interaction angles ($0\leq\theta\leq\pi/2$) correspond to stronger off-diagonal coupling. The steady-state probabilities also increase with interaction angles, as expected from Eq.~(\ref{final_ohmic}). Fig.~\ref{Pop_Ph_D0_diftheta} displays time evolution of the boson number for various interaction angles of (a) $\theta=\pi/4$ and (b) $\theta=\pi/2$. We found that larger interaction angles ($0\leq\theta\leq\pi/2$) lead to more bosons being created during the transition stage, via stronger off-diagonal coupling.

\section{Conclusion}
\label{Conclusions}
In this work, we have studied the intriguing role played by the dissipative environment in LZ dynamics. Following the Dirac-Frenkel time-dependent variational principle, dynamics of the LZ model with diagonal and off-diagonal qubit-bath coupling is probed by employing the multi-D$_2$ {\it Ansatz}, also known as a linear combination of the usual Davydov D$_2$ trial states. Convergence has been ensured in the LZ dynamics calculation by monitoring the multiplicity of the multi-D$_2$ {\it Ansatz}, and results agree with those of other methods. The final transition probabilities in the steady states obtained from our numerical calculations concur with the analytical predictions. To our knowledge, two-stage LZ transitions are found for the first time at the intermediate times for a qubit coupled to a circuit oscillator, or to a dissipative environment that is characterized by the bath spectral densities, thanks to the combined effect of off-diagonal qubit-bath coupling and the tunneling strength. It is revealed in our systematic investigations that larger interaction angle ($0\leq\theta\leq\pi/2$) and spectral densities with larger exponents and coupling strengths lead to longer transition times and greater steady-state probabilities. Finally, our boson dynamics analysis based on the multi-$\rm D_2$ {\it Anstaz} has successfully identified the contribution of specific boson modes to the LZ transitions. A detailed understanding of the mechanism using the Lorentz type spectral density in flux qubit and multi-level transitions is of great interest and awaits further investigations.

\section*{Acknowledgments}
The authors would like to thank Qinghu Chen, Nengji Zhou, Kewei Sun, and Lipeng Chen for useful discussions. Support from the Singapore National Research Foundation through the Competitive Research Programme (CRP) under Project No.~NRF-CRP5-2009-04 and from the Singapore Ministry of Education Academic Research Fund Tier 1 (Grant No. RG106/15) are gratefully acknowledged.

\appendix
\section{The time dependent variational approach for the dissipative Landau-Zener model}
\label{Equations of Motion}
In order to apply the Dirac-Frenkel time-dependent variational principle, we first need to calculate the Lagrangian $L_2$,
\begin{eqnarray}
\label{Lagrangian_detail}
&&L_2=\frac{i}{2}\sum_{i,j}\left(A_{j}^{*}\dot{A}_{i}-\dot{A}_{j}^{\ast}A_{i}+B_{j}^{\ast}\dot{B}_{i}-\dot{B}_{j}^{\ast}B_{i}\right)S_{ji}\nonumber\\ &&+\frac{i}{2}\sum_{i,j}\left(A_{j}^{\ast}A_{i}+B_{j}^{\ast}B_{i}\right)\sum_{q}[\frac{\dot{f}_{jq}^{\ast}f_{jq}+f_{jq}^{\ast}\dot{f}_{jq}}{2}\nonumber\\
&&-\frac{\dot{f}_{iq}f_{iq}^{\ast}+f_{iq}\dot{f}_{iq}^{\ast}}{2}+f_{jq}^{\ast}\dot{f}_{iq}-f_{iq}\dot{f}_{jq}^{\ast}]S_{ji}\nonumber\\
&&-\left\langle D_{2}^{M}\left(t\right)\right|\hat{H}\left|D_{2}^{M}\left(t\right)\right\rangle,
\end{eqnarray}
where the Debye-Waller factor is $S_{ji}=\exp{\sum_{q}\left\{-\left(\left|f_{jq}\right|^{2}+\left|f_{iq}\right|^{2}\right)/2+f_{jq}^{\ast}f_{iq}\right\}}$,
and the last term in Eq.~(\ref{Lagrangian_detail}) can be obtained as
\begin{eqnarray}
&&\left\langle D_{2}^{M}\left(t\right)\right|\hat{H}\left|D_{2}^{M}\left(t\right)\right\rangle\nonumber\\
&&=\frac{vt}{2}\sum_{i,j}\left(A_{j}^{\ast}A_{i}-B_{j}^{\ast}B_{i}\right)S_{ji}+\frac{\Delta}{2}\sum_{i,j}\left(A_{j}^{\ast}B_{i}+B_{j}^{\ast}A_{i}\right)S_{ji}\nonumber\\
&&+\sum_{i,j}\left(A_{j}^{\ast}A_{i}+B_{j}^{\ast}B_{i}\right)\sum_{q}\omega_{q}f_{jq}^{\ast}f_{iq}S_{ji}\nonumber\\	&&+\frac{1}{2}\sum_{i,j}\left(A_{j}^{\ast}A_{i}-B_{j}^{\ast}B_{i}\right)\sum_{q}\gamma_{q}\cos\theta_{q}\left(f_{iq}+f_{jq}^{\ast}\right)S_{ji}\nonumber\\ &&+\frac{1}{2}\sum_{i,j}\left(A_{j}^{\ast}B_{i}+B_{j}^{\ast}A_{i}\right)\sum_{q}\gamma_{q}\cos\theta_{q}\left(f_{iq}+f_{jq}^{\ast}\right)S_{ji}.
\end{eqnarray}

The Dirac-Frenkel variational principle results in equations of motion for $A_i$ and $B_i$,
\begin{eqnarray}
&&-i\sum_{i}\dot{A}{}_{i}S_{ki}\nonumber \\	&&-\frac{i}{2}\sum_{i}A_{i}\sum_{q}\left[-\left(\dot{f}_{iq}f_{iq}^{\ast}+f_{iq}\dot{f}_{iq}^{\ast}\right)+2f_{kq}^{\ast}\dot{f}_{iq}\right]S_{ki}\nonumber \\
&&=-\frac{vt}{2}\sum_{i}A_{i}S_{ki}-\frac{\Delta}{2}\sum_{i}B_{i}S_{ki}-\sum_{i}A_{i}\sum_{q}\omega_{q}f_{kq}^{\ast}f_{iq}S_{ki}\nonumber \\
&&-\frac{1}{2}\sum_{i}A_{i}\sum_{q}\gamma_{q}\cos\theta_{q}\left(f_{iq}+f_{kq}^{\ast}\right)S_{ki}\nonumber\\
&&-\frac{1}{2}\sum_{i}B_{i}\sum_{q}\gamma_{q}\sin\theta_{q}\left(f_{iq}+f_{kq}^{\ast}\right)S_{ki},
\end{eqnarray}
and
\begin{eqnarray}
&&-i\sum_{i}\dot{B}_{i}S_{ki}\nonumber \\ &&-\frac{i}{2}\sum_{i}B_{i}\sum_{q}\left[-\left(\dot{f}_{iq}f_{iq}^{\ast}+f_{iq}\dot{f}_{iq}^{\ast}\right)+2f_{kq}^{\ast}\dot{f}_{iq}\right]S_{ki}\nonumber \\
&&=+\frac{vt}{2}\sum_{i}B_{i}S_{ki}-\frac{\Delta}{2}\sum_{i}A_{i}S_{ki}-\sum_{i}B_{i}\sum_{q}\omega_{q}f_{kq}^{\ast}f_{iq}S_{ki}\nonumber \\ &&+\frac{1}{2}\sum_{i}B_{i}\sum_{q}\gamma_{q}\cos\theta_{q}\left(f_{iq}+f_{kq}^{\ast}\right)S_{ki}\nonumber\\
&&-\frac{1}{2}\sum_{i}A_{i}\sum_{q}\gamma_{q}\sin\theta_{q}\left(f_{iq}+f_{kq}^{\ast}\right)S_{ki}.
\end{eqnarray}

The equations of motion for $f_{iq}$ are
\begin{eqnarray} &&-i\sum_{i}\left[\left(A_{k}^{\ast}\dot{A}_{i}+B_{k}^{\ast}\dot{B}_{i}\right)f_{iq}-\left(A_{k}^{\ast}A_{i}+B_{k}^{\ast}B_{i}\right)\dot{f}_{iq}\right]S_{ki}\nonumber\\	&&-\frac{i}{2}\sum_{i}\left(A_{k}^{\ast}A_{i}+B_{k}^{\ast}B_{i}\right)f_{iq}S_{ki}\nonumber\\
&&\times\sum_{p}\left(2f_{kp}^{\ast}\dot{f}_{ip}-\dot{f}_{ip}f_{ip}^{\ast}-f_{ip}\dot{f}_{ip}^{\ast}\right)\nonumber\\
&&=-\frac{vt}{2}\sum_{i}\left(A_{k}^{\ast}A_{i}-B_{k}^{\ast}B_{i}\right)f_{iq}S_{ki}\nonumber\\
&&-\frac{\Delta}{2}\sum_{i}\left(A_{k}^{\ast}B_{i}+B_{k}^{\ast}A_{i}\right)f_{iq}S_{ki}\nonumber\\	&&-\sum_{i}\left(A_{k}^{\ast}A_{i}+B_{k}^{\ast}B_{i}\right)\left(\omega_{q}+\sum\omega_{p}f_{kp}^{\ast}f_{ip}\right)f_{iq}S_{ki}\nonumber\\	&&-\frac{1}{2}\sum_{i}\left(A_{k}^{\ast}A_{i}-B_{k}^{\ast}B_{i}\right)\gamma_{q}\cos\theta_{q}S_{ki}\nonumber\\
&&-\frac{1}{2}\sum_{i}\left(A_{k}^{\ast}A_{i}-B_{k}^{\ast}B_{i}\right)f_{iq}\sum_{p}\gamma_{p}\cos\theta_{p}\left(f_{ip}+f_{kp}^{\ast}\right)S_{ki}\nonumber\\	&&-\frac{1}{2}\sum_{i}\left(A_{k}^{\ast}B_{i}+B_{k}^{\ast}A_{i}\right)\gamma_{q}\sin\theta_{q}S_{ki}\nonumber\\
&&-\frac{1}{2}\sum_{i}\left(A_{k}^{\ast}B_{i}+B_{k}^{\ast}A_{i}\right)f_{iq}\sum_{p}\gamma_{p}\sin\theta_{p}\left(f_{ip}+f_{kp}^{\ast}\right)S_{ki}.\nonumber\\
\end{eqnarray}

It should be noted that the main results of this work are calculated from the above equations of motion. The equations of motion are solved numerically by means of the fourth-order Runge-Kutta method.
In this work, the qubit is assumed to initially occupy the state $\left|\uparrow\right\rangle$, i.e., $A_1(0) = 1$, $B_1(0) = 0$, and $A_i(0) = B_i(0) = 0 (i\neq1)$. The initial bosonic displacement is set to zero ($f_{iq}(t\rightarrow-\infty)=0$), though the LZ transitions have been demonstrated to depend also on various types of initial coherent superposition states \cite{keeling_2008, sun_2012}.

\section{Convergence test of Landau-Zener dynamics for the qubit coupled to a bath of quantum harmonic oscillators}
\label{convergence test}

\begin{figure}[tbp]
\centering
\includegraphics[scale=0.44]{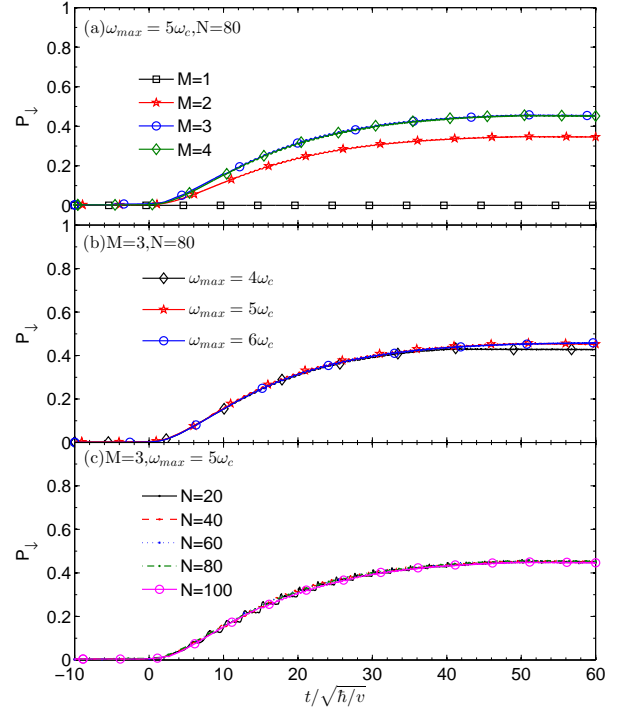}
\caption{Time evolution of transition probability calculated by the multi-$D_2$ {\it Ansatz}. Tested parameters are (a) number of multiplicity $M$, (b) maximum spectrum band frequencies $\omega_{max}$, and (c) number of oscillator modes $N$. Other parameters are $\Delta=0$, $\alpha=0.002$, $s=1$, and $\omega_c=10\sqrt{v/\hbar}$.}
\label{SI_dif_M_wmax_N}
\end{figure}

We have performed convergence tests using the multi-D$_2$ {\it Ansatz} for the qubit that is coupled to a bath of harmonic oscillators. As shown in Fig.~\ref{SI_dif_M_wmax_N}(a), (b), and (c), we have studied the effects of the multiplicity $M$, maximum cutoff frequency $\omega_{max}$, and number of modes $N$ on numerical calculations, respectively. As shown Fig.~\ref{SI_dif_M_wmax_N}(a), multiplicity $M$ of $1$, $2$, $3$, and $4$ are adopted in the calculations. It is found that converged results can be obtained using $M=3$ for the studied multiple-mode scenario, which also contains low-frequency bath oscillators. In contrast, for the single low-frequency mode case, a much larger multiplicity of $M=7$ is required for the convergence as shown in Fig.~\ref{SI_converge_Omeaga}(a). In the following we briefly explain why a large multiplicity is not necessary in the presence of multiple low-frequency modes. As for Fig.~\ref{SI_converge_Omeaga}, the convergence test is performed for a single oscillator case. Before $t=\sqrt{\hbar/v}$ we have already obtained converged results using $M=3$ in the case of $\omega=0.1\sqrt{v/\hbar}$. Around $t=\sqrt{\hbar/v}$ the LZ transition of $\omega=\sqrt{v/\hbar}$  appears much faster than that of $\omega=0.1\sqrt{v/\hbar}$ before the onset of the steady state. This indicates that a small multiplicity of $M=3$ is sufficient to get accurate results if both frequencies of $\omega=0.1\sqrt{v/\hbar}$ and $\omega=\sqrt{v/\hbar}$ are included. As for Fig.~\ref{SI_dif_M_wmax_N} (a), the convergence test is performed with respect to multiple harmonic oscillators, which contain both frequencies of $\omega=0.1\sqrt{v/\hbar}$ and $\omega=\sqrt{v/\hbar}$. Therefore, the multiplicity of $M=3$ is satisfactory to provide accurate LZ dynamics. Meanwhile, the steady-state probability of $M=3$ also agrees with the analytical prediction\cite{saito_2007}. As presented in Fig.~\ref{SI_dif_M_wmax_N}(b), maximum cutoff frequency $\omega_{max}$ of $4\omega_c$, $5\omega_c$, and $6\omega_c$ are used with $\omega_c=10\sqrt{v/\hbar}$. It can be found that $\omega_{max}=5\omega_c$ is sufficient to get converged results. Fig.~\ref{SI_dif_M_wmax_N}(c) presents LZ dynamics using the number of oscillator modes $N$ of $20$, $40$, $60$, $80$, and $100$. The roughness of the curves is found to be smaller as the number of modes becomes larger. After the careful convergence tests, the well tested parameters have been applied in the numerical calculations in this work.

\end{document}